\documentclass[apex]{jjap3}

\usepackage[T1]{fontenc}
\usepackage[latin1]{inputenc}
\usepackage{amsmath}
\usepackage{graphicx}
\usepackage{amssymb}

\usepackage{color} 
\usepackage{graphicx}
\usepackage{bm}
\usepackage{amsmath}

\usepackage{gt15apex}

\begin{document}
\title{
Efficient stopping of current-driven domain wall using a local  Rashba field\\ 
}
\author{{Gen Tatara},{Henri Saarikoski},{Chiharu Mitsumata}$^{1}$}

\inst{RIKEN Center for Emergent Matter Science (CEMS), 2-1 Hirosawa, Wako, Saitama, 351-0198 Japan\\
$^{1}$ National Institute for Materials Science,  1-2-1 Sengen, Tsukuba, Ibaraki, 305-0047 Japan}

 \abst{
We theoretically show  that a locally-embedded Rashba interaction acts as a strong pinning center for current-driven domain walls  and demonstrate efficient capturing and depinning of the wall using a weak Rashba interaction of the order of 0.01 eV\AA.
Our discovery is expected to be useful for the highly reliable control of domain walls in racetrack memories. 
}

\date{\today}


\maketitle

Magnetic memories operated fully electrically are promising for fast and high-density memories. 
Switching of the magnetization of a thin ferromagnetic layer by applying an electric current has been accomplished using the spin-transfer effect \cite{Berger96,Slonczewski96}. 
Driving a ferromagnetic domain wall by current pulse has been also achieved.
Most in-plane magnetic anisotropy systems are in the extrinsic pinning regime, where the wall motion is  driven by a non-adiabatic torque \cite{Zhang04,Thiaville05}, while  the spin-transfer torque in the intrinsic pinning regime \cite{TK04} has been reported in a perpendicular magnetization material \cite{Koyama11}.
Recently, several possibilities for using multilayers for fast domain wall motion have been proposed \cite{Emori13,YangParkin15,Saarikoski14,Lepadatu16}. 
For ultrahigh density memories, use of a sequence of domain walls on a patterned wire (``race track'') controlled by electric current, called a racetrack memory, has been proposed \cite{Parkin15}.
For memory applications of such multi-domain wall devices, techniques to stop a moving wall at an intended position precisely and without delay are essential. 
An artificial pinning site has been proposed for stopping the wall \cite{Parkin08}; however, efficient and reliable stopping is difficult because of difficulty in fabricating well-controlled and uniform pinning centers. 
Moreover, fast stopping requires a strong pinning potential, which necessitates a large current density for depinning.

In this paper, we propose a highly efficient and reliable mechanism to stop a moving domain wall using a locally-embedded Rashba spin-orbit interaction.
The  Rashba interaction generates a strong effective magnetic field when an electric  current is injected \cite{Obata08,Manchon09} . 
This effective field leads to strong pinning of a moving wall at the Rashba region if the applied current density is below the capturing threshold $j_{\rm cap}$.
Moving the wall from the pinning center is performed by applying a higher current pulse, above the depinning threshold $j_{\rm dep}$.
The Rashba pinning is highly reliable because introducing a Rashba interaction by attaching a small thin layer of heavy metals in a controlled manner is easy using the present technology.
The present mechanism also has an advantage of a lower energy consumption compared with geometrical pinning. 
The fact that the capturing threshold $j_{\rm cap}$ is lower than $j_{\rm dep}$ indicates that the energy required to shift the wall positions over a distance of multiple pinning sites is much lower than that in the case of  the geometrical pinning mechanism.

\begin{figure}[t]
\begin{center}
 \includegraphics[width=0.4\hsize]{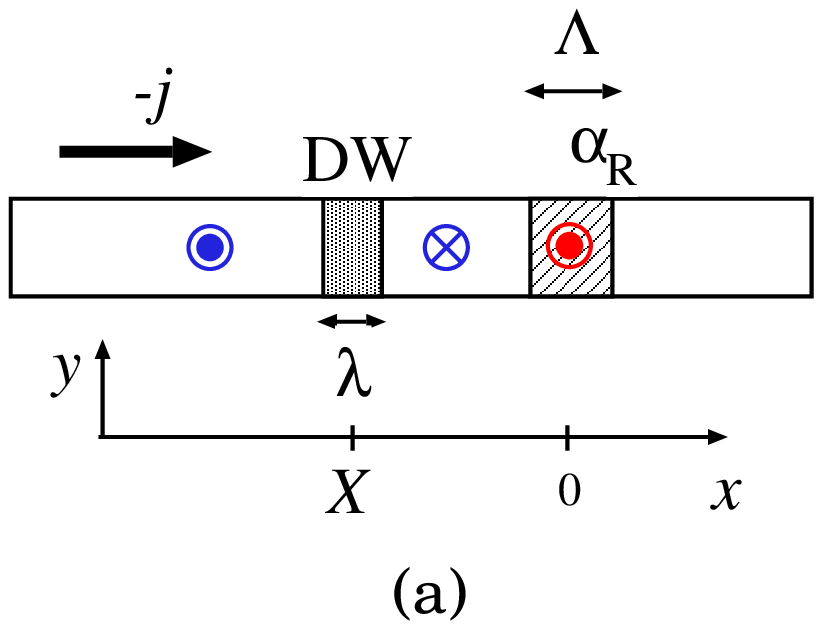} 
 \includegraphics[width=0.4\hsize]{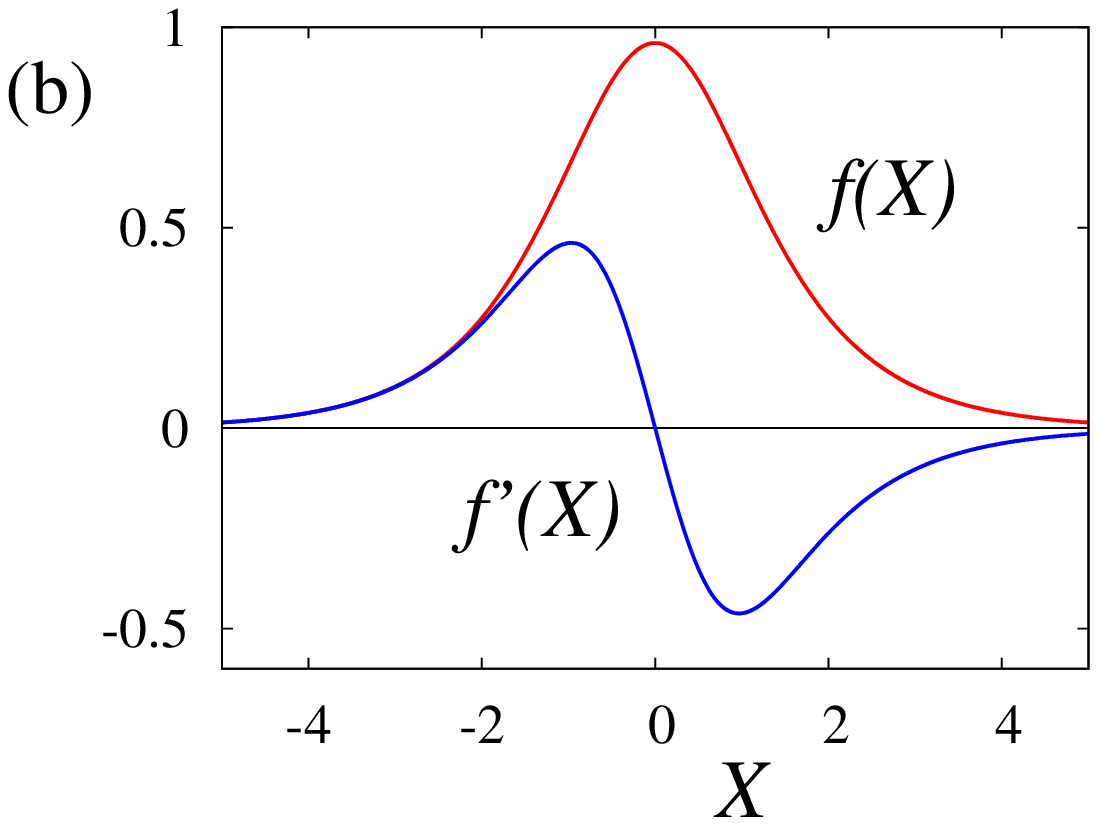} 
\end{center}
 \caption{ 
(a) Setting of the system.  A domain wall in a ferromagnetic wire  along the $x$ axis is driven by a steady current $j$ along $-\hat{x}$ direction. The Rashba interaction, represented by a vector $\alphaRv$, is embedded in the shaded region near $x=0$.   From the symmetry, $\alphaRv$ is along $z$ axis (perpendicular to the wire) and produce a local strong effective magnetic field along $-y$ direction when a current is applied. 
 (b) Plots of $f(X)$ and $f'(X)$ representing profiles of pinning torque and force, respectively,  for $\lambda=1$ and $\Lambda=1$.
 \label{FIGgeometry}}
\end{figure}

The system we consider to demonstrate the Rashba pinning effect is simple;  
it includes a ferromagnetic wire with Rashba interaction locally embedded by attaching a small thin film of heavy metals (Fig. \ref{FIGgeometry}(a)).
The Rashba field, represented by a vector $\alphaRv$, is perpendicular to the wire plane.
The current applied along the wire then generates an in-plane effective magnetic field orthogonal to the wire.
The $z$ axis is selected along $\alphaRv$, while the $x$ axis is selected along the wire.
The wall we consider is a Bloch wall, with the magnetic easy axis along the $z$ direction and the wall plane at equilibrium lying in the $yz$ plane. 
The wall structure was stabilized by the current when the wall was in the Rashba region because of  the generated magnetic field along the $y$ direction.
The Hamiltonian of localized spin $\Sv$  is given by 
\begin{align}
 H&= \int \frac{d^3 r}{a^3}\lt[\frac{J}{2}(\nabla\Sv)^2-\frac{K}{2}(S_z)^2+\frac{\Kp}{2}(S_x)^2 
 +E_{{\rm R},S}\rt],
\end{align}
where $J$, $K$, $\Kp$ and $a$ are the exchange energy, easy and hard axis anisotropy energies,  and the lattice constant, respectively. The last term $E_{{\rm R},S}$ describes the influence of effective magnetic field due to the Rashba interaction.
The wall configuration centered at $x=X$ is represented as \cite{TKS_PR08}
\begin{align}
 S_x&=\frac{S \cos\phi}{\cosh \frac{x-X}{\lambda}}, 
 S_y=\frac{S \sin\phi}{\cosh \frac{x-X}{\lambda}}, 
 S_z= S \tanh\frac{x-X}{\lambda}  , \label{DWsol}
\end{align}
where $\lambda=\sqrt{J/K}$ is the width of the wall, $\phi$ is the angle of the wall plane and  $S$ is the magnitude of localized spin.

The  Rashba interaction embedded in the region $-\frac{\Lambda}{2}<x<\frac{\Lambda}{2}$ is represented by a Hamiltonian 
\begin{align}
 H_{\rm R} = -\frac{1}{\hbar}[\alphaRv\cdot(\pv\times\sigmav)] \theta_{\rm R}(x),
\end{align}
where $\pv$ is the electron's momentum and $\sigmav$ is a vector of Pauli matrices and 
$ \theta_{\rm R}(x)\equiv (1-\theta({\Lambda}/{2}))\theta(-{\Lambda}/{2})$, where $\theta(x)=1$ for $x\geq 0$ and  $\theta(x)=0$ for $x<0$. 
We assumed that the Rashba field applies homogeneously in the region  $-\frac{\Lambda}{2}<x<\frac{\Lambda}{2}$.
When the current density $\jv$ is applied to the wire, conduction electron spin in the Rashba region experiences a magnetic field of 
$\Bv_{\rm e}=\frac{ma^3}{-e\hbar^2\gamma}\alphaRv\times\jv$
where $m$ is the electron mass, $e(<0)$ is the electron charge,  and $\gamma(=\frac{|e|}{m})$ is  the gyromagnetic ratio.
In the case of a strong $sd$ exchange interaction between the localized and electron spins (adiabatic limit), the field acting on the localized spins is $\Bv_{\rm e}$ but with $j$ replaced by spin current $\js\equiv Pj$, namely
\begin{align} 
\Bv_{\rm R}=-B_{\rm R}\theta_{\rm R}(x)\hat{\yv},
\end{align}
where $\hat{\yv}$ is the unit vector along the $y$ axis and 
$B_{\rm R}\equiv  \frac{ma^3P}{-e\hbar^2\gamma S}\alphaR j$, where we chose positive $j$ as along $-x$ direction.

It was theoretically pointed out that the Rashba interaction induces besides the field $\Bv_{\rm R}$ an effective magnetic field along the direction $\Bv_{\rm R}\times\nv$, where $\nv$ denotes the localized spin direction \cite{WangRashba12,KimRashba12}.
This field, a perpendicular field, turned out to be smaller than $\Bv_{\rm R}$ by a factor of $0.01$-$0.2$ in dirty metals \cite{WangRashba12}. 
Moreover, the perpendicular field would not affect much the pinning effect due to $\Bv_{\rm R}$ in the present case, as the larger field   $\Bv_{\rm R}$ tends to pin the wall by pointing $\nv$ inside the wall along $\Bv_{\rm R}$, resulting in small $\Bv_{\rm R}\times\nv$.
In the following calculation, we therefore neglect the effect of the perpendicular effective field, and focus on the dominant pinning effect by the field    $\Bv_{\rm R}$.

The additional energy of the wall arising from the Rashba-induced field $\Bv_{\rm R}$ is  given by 
\begin{align}
 E_{{\rm R},S}
 = \frac{\Nw ma^3P}{2e\hbar }\alphaR j_{\rm R}  f(X) \sin\phi,
\end{align}
where $A$ is the cross section of the wire and $\Nw=\frac{2\lambda A}{a^3}$ is the number of spins in the wall and 
\begin{align}
f(X) \equiv & \int_{-\frac{\Lambda}{2}-X}^{\frac{\Lambda}{2}-X} \frac{dz}{\lambda} 
    \frac{1}{\cosh\frac{z}{\lambda}}
    =
2\lt[\tan^{-1} e^{ (\frac{\Lambda}{2}-X)/\lambda }
- \tan^{-1} e^{ (-\frac{\Lambda}{2}-X)/\lambda}\rt].
\end{align}
The force on the wall due to the Rashba field is given by 
\begin{align}
- \frac{\delta E_{\rm R}}{\delta X} 
=& 
\frac{\Nw ma^3}{-2e\hbar  \lambda}\alphaR j_{\rm R}  \lambda f'(X) \sin\phi \nnr
\lambda f'(X) =& \frac{1}{\cosh \frac{\frac{\Lambda}{2}+X}{\lambda} } - \frac{1}{\cosh \frac{\frac{\Lambda}{2}-X}{\lambda} } .
\end{align}
The behaviors of functions $f$ and $f'$ are plotted in Fig. \ref{FIGgeometry}(b).

Including Gilibert damping $\alpha$ and the $\beta$ (nonadiabaticity) term, the equation of motions for the wall are \cite{TKS_PR08} 
\begin{align}
 \dot{\phi}+\alpha\frac{\dot{X}}{\lambda} =&
 P \jtil\lt[ \frac{\beta}{\lambda} - \tilde{\alphaR} f'(X) \sin\phi \rt] \nnr
 \dot{X}-\alpha\lambda \dot{\phi} =& 
 -\vc\sin2\phi + P\jtil \lt[1 +  \tilde{\alphaR} f(X) \cos\phi \rt] .
 \label{eqs}
\end{align}
Here $\vc\equiv \frac{K_{\perp}\lambda S}{2\hbar}$, 
\begin{align}
\tilde{\alphaR} \equiv &
  \frac{m\lambda}{\hbar^2 } \alphaR ,  \;\;\; 
  \jtil \equiv    \frac{a^3}{2eS}j,
\end{align}
where $\ef$ and $\kf$ are the Fermi energy and Fermi wavelength, respectively.
As $e<0$, $\jtil$ is opposite to $j$ and the wall moves toward the direction of $\jtil$. 
For positive $\jtil$, the wall favors the configuration $\phi\simeq -\frac{\pi}{2}$ because of the Rashba effect. 

\begin{figure}[t]
\centering
 \includegraphics[width=0.4\hsize]{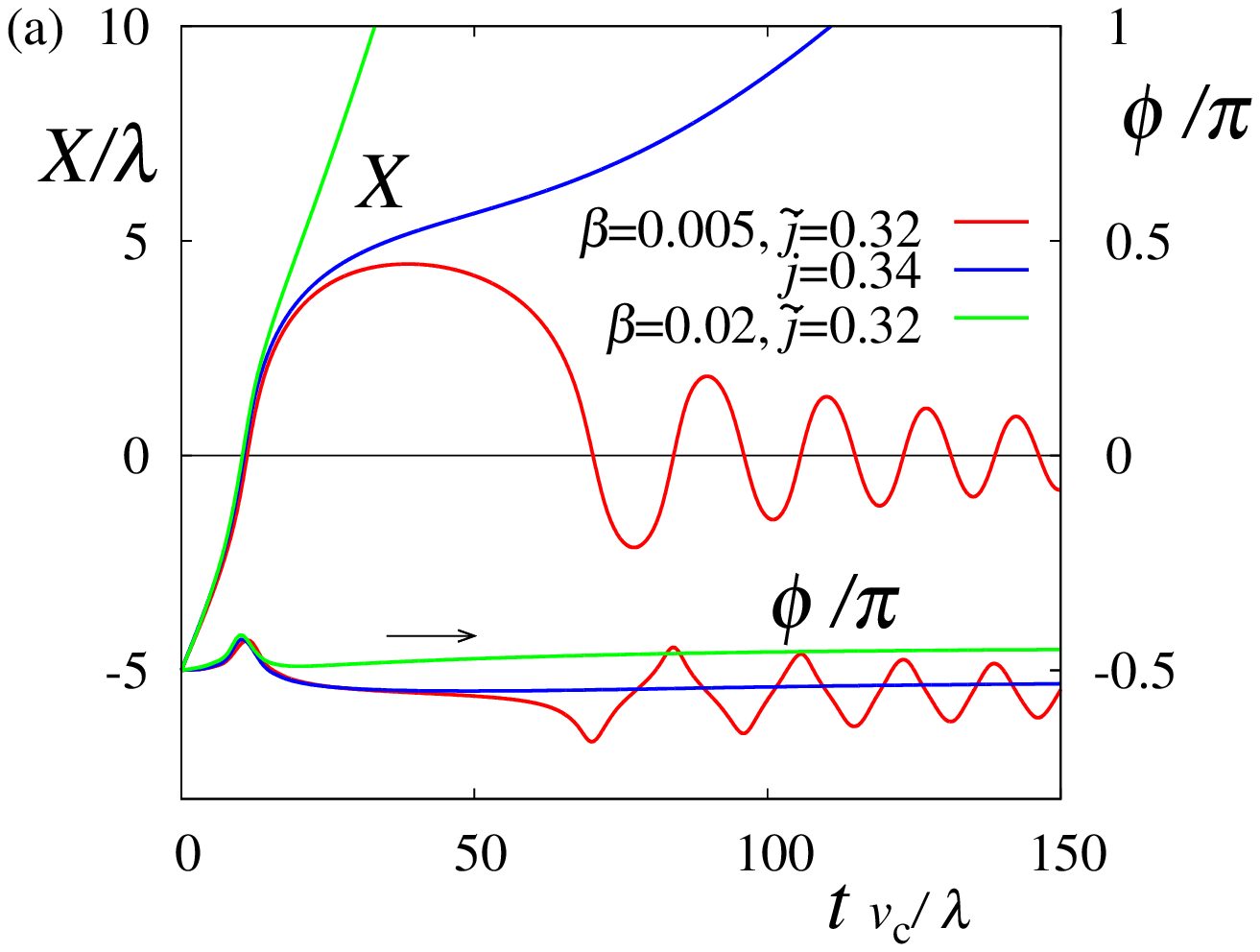}
 \includegraphics[width=0.5\hsize]{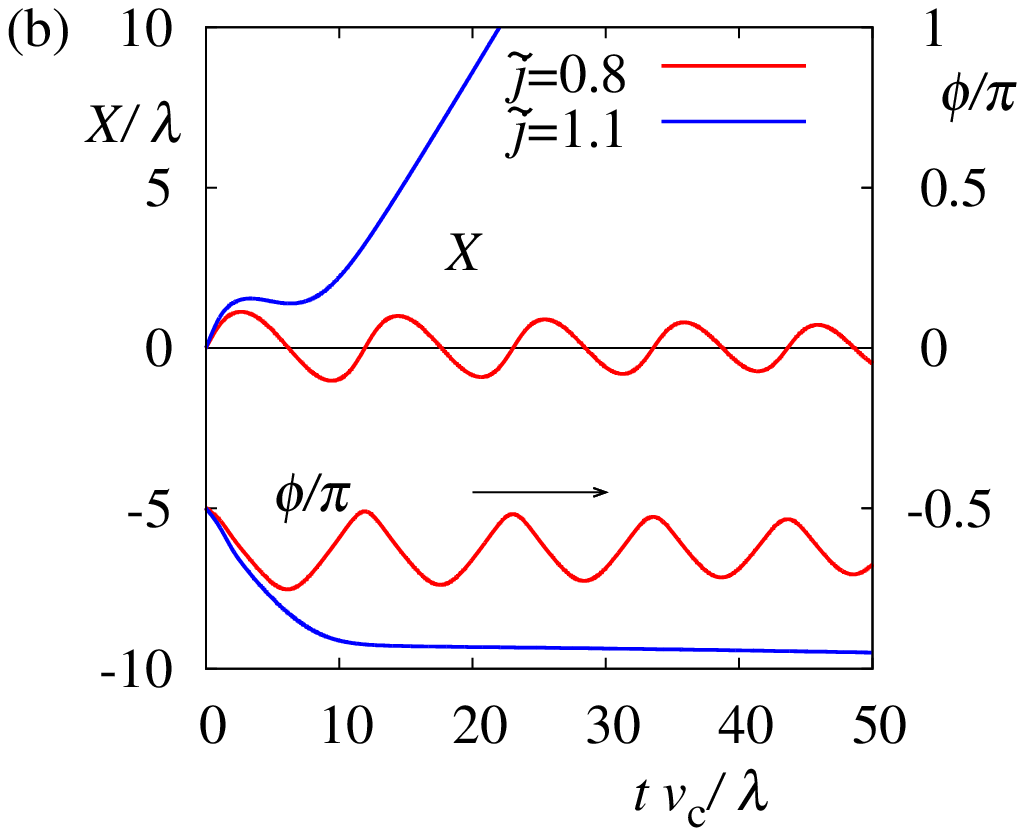}
 \caption{ Plots of wall position $X$ and $\phi$ as function of time for the case of $\tilde{\alphaR}=0.5$ with $\alpha=0.01$ and $\Lambda=1$.
 Plot (a) is for the initial condition $X/\lambda=-5$ and $\phi=-\frac{\pi}{2}$  at $t=0$. 
 The normalized current $\jtil$ below and above the threshold value $j_{\rm cap}(\sim0.33)$ are shown for the case of $\beta=0.005$. 
 Plot (b) shows the depinning behavior from $X=0$ and $\phi=-\frac{\pi}{2}$  at $t=0$ for $\jtil$ below and above the threshold value $j_{\rm dep}(\sim0.9)$.
 \label{FIGXphit}}
\end{figure}
\begin{figure}[th]
\centering
 \includegraphics[width=0.8\hsize]{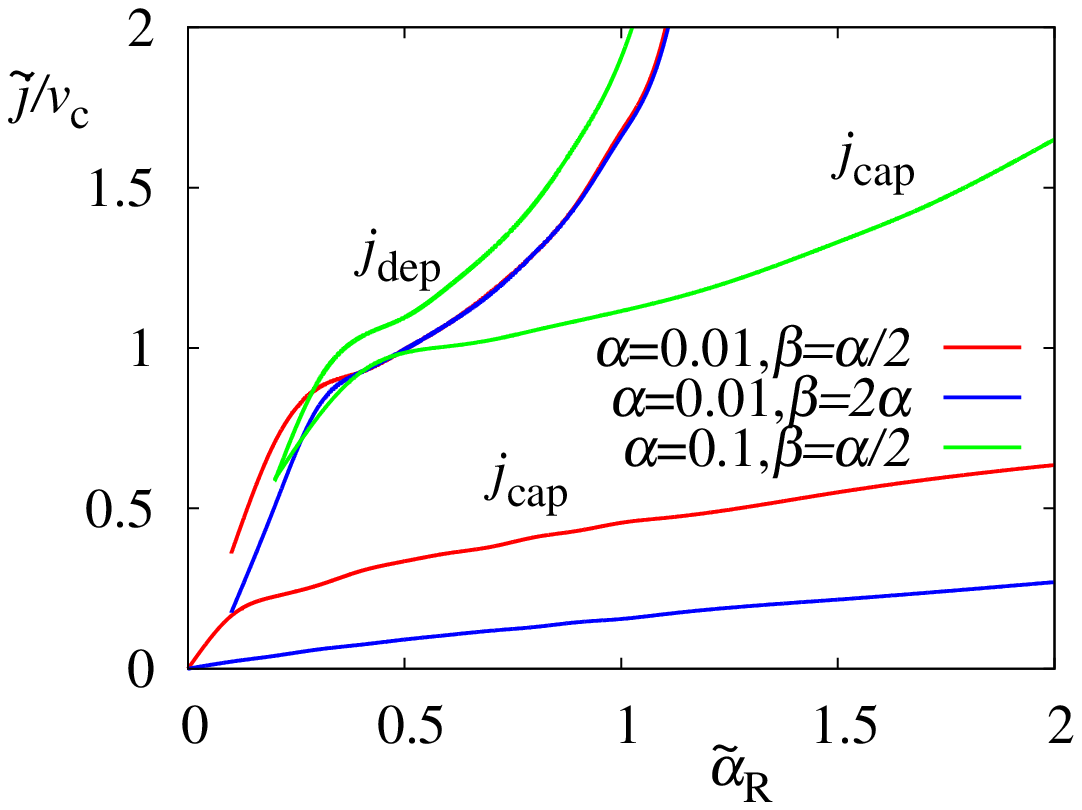}
 \caption{ Plot of threshold current densities for capturing the wall, $j_{\rm cap}$, and for depinning the wall from the Rashba pinning potential, $j_{\rm dep}$, both normalized by $\vc$ plotted as functions of $\tilde{\alphaR}$. 
 Below $j_{\rm cap}$, the wall is captured by the Rashba pinning potential. When a current above $j_{\rm dep}$ is applied, the captured wall is depinned.
 \label{FIGtrananddepin}}
 \end{figure}

Typical solutions of the equation of motions are shown in Fig. \ref{FIGXphit}.
Figure \ref{FIGXphit}(a) shows the capturing dynamics of the wall under current driven from the initial position outside the Rashba regime ($X/\lambda=-5$).
We found that there is a threshold current density $j_{\rm cap}$ below which the wall is captured by the Rashba pinning region. 
The capturing threshold is plotted as a function of $\alphaR$ in Fig. \ref{FIGtrananddepin}, and is an order of magnitude lower than the intrinsic pinning threshold
$\jtil_{\rm i}= \frac{\vc}{P}$ \cite{TKS_PR08} if $\alphaR\simeq O(1)$. 
An important observation here is that the Rashba field blocks  the wall motion with a realistic value of current $\jtil\lesssim O(1)$ if $\tilde{\alphaR}$ is of the order of unity, which corresponds to a rather weak value of $\alphaR=0.01$ eV\AA.
Thus, the weak Rashba interaction is enough for practical devices.
The  capturing threshold  depends on $\beta$ since the wall speed when it enters the Rashba region depends on $\beta$. 

\begin{figure}[t]
\centering
 \includegraphics[width=0.4\hsize]{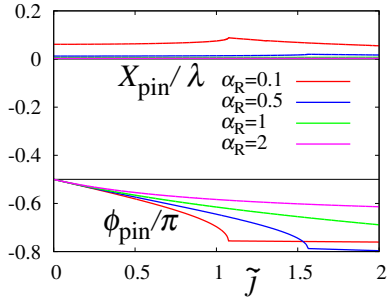}
 \caption{ Plots of equilibrium pinned values of $X_{\rm pin}$ and $\phi_{\rm pin}$ as function of $\jtil$ for the case of $\alpha=0.01$ and $\beta=0.005$.  
 \label{FIGXphieq}}
\end{figure}

Let us analyze the pinning mechanism. 
We focus on the  wall configuration near $X\sim0$ and $\phi=-\frac{\pi}{2}$ and write 
$f'(X)\simeq -f_1 X/\lambda^2  $.
The equilibrium pinned configuration at finite $\jtil$ is then given by
\begin{align}
 X_{\rm pin} &=\frac{\beta \lambda}{f_1 \tilde{\alphaR} } \nnr
 \delta \phi_{\rm pin} &\equiv \phi_{\rm pin}+\frac{\pi}{2} =-\frac{\jtil}{\frac{2\vc}{P}+f_0 \tilde{\alphaR} \jtil},
\end{align}
where $f_0\equiv f(0)$.
Since $\beta\ll1$, we see that the captured position of the wall is very close to the center of the Rashba region as far as $\tilde{\alphaR}\gtrsim O(1)$.
Deviation of the angle, $\delta \phi$, is small only when either $\jtil\ll1$ or $\jtil \gg \frac{\vc}{P\tilde{\alphaR}}$. 
The equilibrium pinned configuration obtained from Eq. (\ref{eqs})  is plotted in Fig. \ref{FIGXphieq}. 
$\phi_{\rm pin}$ is insensitive to $\beta$ in contrast to $X_{\rm pin}$.  

When the deviation $\delta\phi_{\rm pin}$ is not small, the wall moves backward when the current is stopped. In fact, the wall speed and $\dot{\phi}$ in the absence of current satisfy 
$\dot{X}=-\frac{\dot{\phi}}{\alpha}\lambda$, which means that the wall moves over a substantial distance of 
$\delta X= -\frac{\delta{\phi}_{\rm pin}}{\alpha}\lambda$ when the phase $\phi$ reduces to the equilibrium value of $-\frac{\pi}{2}$ at $j=0$.
In reality this would not be serious for devices, since the backward motion is removed simply by use of a smooth cut of the current as is understood from Fig. \ref{FIGXphieq}.

To see the capturing dynamics, let us expand the equation of motion with respect to $X$ and $\delta\phi$.
Neglecting small quantities of the order of $\alpha^2$ and $\alpha\beta$, we obtain 
\begin{align}
 \dot{X} =&  P\jtil+\delta\phi [2\vc + Pf_0\tilde{\alphaR} \jtil] -\alpha \tilde{\alphaR} f_1 P\jtil \frac{X}{\lambda} 
     \nnr
 \lambda \dot{\delta\phi}=&
 P \jtil(\beta-\alpha)  -\alpha \delta\phi [2\vc + Pf_0\tilde{\alphaR} \jtil] -  \tilde{\alphaR} f_1 P\jtil \frac{X}{\lambda} 
 \label{eqssmall}
\end{align}
The second equation indicates that when the wall enters the region of $X>0$ from the $X<0$ region, $\delta\phi$ starts to be negative because $\dot{\delta\phi}\simeq -  \tilde{\alphaR} f_1 P\jtil \frac{X}{\lambda} <0$,  resulting  in slow down of the wall because of the first equation. 
The wall is captured if this deceleration is strong for the initial wall velocity proportional to $\jtil$.
The correlation between $\dot{X}$ and $\delta\phi$ is clearly seen in the case of $\jtil=0.32$ of Fig. \ref{FIGXphit}(a).

The time needed for capturing the wall, $\tau_{\rm cap}$, is an important parameter for devices.
It is estimated based on the linearlized equations, (\ref{eqssmall}), which reduces to a second-order differential equation of 
\begin{align}
 \ddot{\delta\phi} + \dot{\delta\phi}\frac{\alpha}{\lambda}  [2\vc + \overline{f}\tilde{\alphaR} P\jtil]
  +\delta\phi \frac{f_1}{\lambda^2}\tilde{\alphaR}  P\jtil [2\vc + {f_0}\tilde{\alphaR} P\jtil]
  =& - \frac{f_1}{\lambda^2}\tilde{\alphaR} (P\jtil)^2,
 \label{eqssmall2}
\end{align}
where $\overline{f}\equiv \frac{1}{2}(f_0+f_1)$.
The time of capturing is given by the imaginary part of the angular frequency $\omega$ of the solution, $\delta\phi\propto e^{-i\omega t}$, as 
\begin{align}
\tau_{\rm cap} &= -\frac{1}{\Im \omega}
 = \frac{\lambda}{\alpha \vc}\lt(1+ \frac{P\jtil}{\vc}\overline{f}\tilde{\alphaR}\rt)^{-1}. 
\end{align}

We choose the hard-axis energy as $\Kp\sim 8\times 10^{-26}$J per site, which would be reasonable in perpendicular media like CoNi \cite{Koyama11}.
The wall thickness is reported to be less than 10 nm.
Choosing $\lambda=5$nm, the intrinsic pinning threshold, $j_{\rm i}\equiv \frac{e S^2}{Pa^3\hbar}\Kp\lambda$ \cite{TK04} is of the order of $10^{11}$A/m$^2$, consistent with experimental results \cite{Koyama11}. 
 In this case, $\vc$ is about 2 m/s and the capturing time is 
$\tau_{\rm cap}=250$ ns if $\alpha=0.01$ and if $\frac{\jtil}{\vc}\tilde{\alphaR}\ll1$.
This seems rather slow, so systems having either larger $\Kp$ or a larger Gilbert damping parameter $\alpha$ are favorable for fast devices. 
Extrinsic enhancement of  the Gilbert damping parameter as proposed in Ref. \cite{Mitsumata11} may also be useful.

We have seen that the stopping the wall at the Rashba pinning potential is realized by applying a current density below $j_{\rm cap}$. 
To set the wall in motion again by applying a new current pulse, there is another threshold value for depinning, since a pinning potential is generated as soon as  the current is injected. 
The depinning threshold current density is  numerically calculated by considering the initial condition of $X=0$ and $\phi=-\frac{\pi}{2}$. The wall dynamics is shown in Fig. \ref{FIGXphit}(b), and the depinning threshold is plotted in Fig. \ref{FIGtrananddepin}.
We see that the $\jtil_{\rm dep}/\vc$ is larger than unity for $\tilde{\alphaR}\gtrsim 0.6$, while it decreases for small $\tilde{\alphaR}$.  
For device operation,  two distinct values of current density, one lower than $j_{\rm cap}$ and the other larger than $j_{\rm dep}$, are used for moving and stopping the wall at an intended position. 
Materials with $0.1 \lesssim \tilde{\alphaR} \lesssim1$ would be suitable for such operations.
To shift the wall position, the initial magnitude of the pulse needs to be larger than $j_{\rm dep}$, but required current to overcome unnecessary pinning sites is lower than $j_{\rm dep}$ (and above $j_{\rm cap}$).
The current needs to be  vanish smoothly at the intended pinning site to avoid a backward motion.

\begin{figure}[t]
\begin{center}
 \includegraphics[width=0.4\hsize]{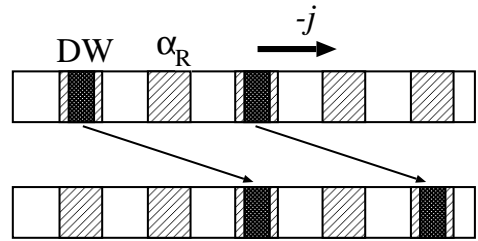} 
\end{center}
 \caption{ Schematic figure showing a shift memory operation based on the Rashba pinning effect.
 The top configuration corresponds to the information 10100, where 1 and 0 correspond to pinning site (shown by shaded areas) with and without a domain wall (shown by black). 
 To shift the wall position to the bottom configuration indicating 00101 is carried out by a current pulse with average amplitude smaller than the depinning threshold (but above $j_{\rm cap}$).
 \label{FIGdws}}
\end{figure}

The results we have found are striking since they show that even a weak Rashba interaction of the order of $\alphaR=0.01$ eV\AA\ is sufficient for stopping the wall when locally introduced .
Instead, strong Rashba interaction like $\alphaR=3$ eV\AA\  as realized on Bi/Ag \cite{Ast07}  is not suitable because $\tilde{\alphaR}$ then is about 300 and the wall is too strongly pinned. 
From this aspect, we believe there are many candidate systems for the present Rashba pinning device. 
Our discovery is expected to be useful for realizing domain wall based shift memories like a racetrack memory.

Our results also suggest that heavy atom impurities may cause strong pinning by modifying in-plane magnetic anisotropy energy as a result of the Rashba-induced magnetic field. 
This possibility seems consistent with the fact that extrinsic pinning effect is dominant in in-plane easy axis anisotropy materials.

\acknowledgements
This work was supported by a 
Grant-in-Aid for Scientific Research on Innovative Areas (Grant No.26103006) from The Ministry of Education, Culture, Sports, Science and Technology (MEXT), Japan and  a Grant-in-Aid for Scientific Research (C) (Grant No. 26390014) from the Japan Society for the Promotion of Science.  


\end{document}